\documentclass[12pt]{iopart}
\usepackage{graphicx}
\usepackage{enumerate}
\graphicspath{{Figurer/}}

\newcommand{\avec}{\textbf{a}}
\newcommand{\Fvec}{\textbf{F}}
\newcommand{\gvec}{\textbf{g}}
\newcommand{\rvec}{\textbf{r}}
\newcommand{\be}{\beta}
\newcommand{\D}{\Delta}
\newcommand{\en}{\epsilon}
\newcommand{\ta}{\theta}
\newcommand{\ti}{\theta_i}
\newcommand{\Dc}{\mathcal{D}}
\newcommand{\F}{\mathcal{F}}
\newcommand{\I}{\mathcal{I}}
\newcommand{\M}{\mathcal{M}}
\newcommand{\Q}{\mathcal{Q}}
\newcommand{\trho}{\widetilde\rho}

\newcommand\B[1]{\textbf{#1}}

\begin{document}

\title[Swinging counterweight trebuchet]
{The swinging counterweight trebuchet\\
On scaling and optimization}

\author{E Horsdal}
\address{Department of Physics and Astronomy, Aarhus University,
DK-8000 Aarhus C, Denmark}
\ead{horsdal@phys.au.dk}
\begin{abstract}
The swinging counterweight trebuchet was at its full blossom in the 
Late Middle Ages an artillery piece of great importance that could destroy 
the defenses of castles under siege by bombardment with stones from 
considerable distances.
The exact dimensions of these engines are unfortunately unknown, but would 
have been valuable for an assessment of the true significance of the weapon.
Presumed dimensions derived from interpreted historical accounts must be 
tested and examined critically.
This has been done by full-scale experimental reconstructions or by~\textit{ad hoc} 
theoretical estimates.
A more systematic approach is made possible by the present theory, which is based
on a mathematical quality function for the trebuchet.
This function can be maximized and the maximizing design parameters are by
definition the optimal parameters.
A simple analytical expression for these optimal parameters is given.
\end{abstract}

%
\section{Introduction}\label{sec:int}
Powered by gravity alone, the swinging counterweight trebuchet gradually became 
the most important artillery piece within the arsenals of several civilizations 
beginning around the year~1200 or earlier and lasting for more than 200 years 
into the era of gunpowder weaponry beginning during the late part of the 
Hundred Years' War~\cite{ref:PVH,ref:Chev_et_al,ref:Chev,ref:MSF2}.
Improvements of the weapon over the period were based entirely on intuition, 
craftsmanship and trial and error~\cite{ref:PVH}, because the physical and 
mathematical sciences were not sufficiently developed to be of much help.
It is interesting nowadays to investigate the full theoretical potential of the 
weapon by using contemporary methods even though it is difficult to decide how 
much this unknown potential was unfolded historically.

A trebuchet with hinged counterweight is composed of three pendulums.
A single pendulum is a relatively simple dynamical system, but three acting 
together and not limited to small swings constitute a quite complex 
problem that is difficult to comprehend, and tuning it for a specific purpose
is a difficult task.
It is not a good strategy to focus on individual parts, one after the other,
because the motion of one part depends intricately on the motion of the two 
others.
Instead, all parameters must be adjusted simultaneously and it requires 
a complete solution to the dynamical problem.

A discussion of good design presupposes agreement on what it entails, 
and the formulation must be sufficiently precise that it can be expressed 
mathematically as a quality function~$\Q$ that can be maximized.
We suggest that~$\Q$ balances the opposing desires for having both a powerful 
and a light and durable weapon that is easily built, moved and operated.
Thus, a trebuchet of good construction must have great destructive potential without 
resorting to excessive overall size or weight.
The function~$\Q$ must therefore depend on range relative to size, on projectile mass
relative to counterweight mass, on efficiency of transforming energy from potential 
to kinetic, and on internal reaction forces that wear down the machine.
Different functions~$\Q$ can be constructed, and any design 
that maximizes a quality function can be considered optimal, so this 
property is not unique, but
different optimized design turn out to be quite similar, so a
clear picture emerges after all.

Capacities of historical trebuchets are most often specifyed by range and 
mass of the projectile, which are~$R$ and~$m$, 
respectively~\cite{ref:MSF1,ref:Galan}. 
Here, we replace~$m$ by the energy~$T$ delivered to the target, 
because the pair~$R$ and~$T$ is a better starting point for analysis
and aligns with contemporary assessment of artillery. 
Except in a few specific cases, it is also assumed for simplicity that aerodynamic 
and sliding friction losses can be ignored and that the target is at the same 
elevation as the base of the engine.

The many adjustable parameters of a trebuchet opens a huge space of conceivable
design which is difficult to navigate without guidelines.
Craftsmanship and experience from the battlefield drove the historical 
development and good trebuchets were most likely built in many cases,
but early written information to document this is 
unfortunately sparse and the available pictorial renditions are difficult to interpret.
Contemporary studies of sieges, where trebuchets were deployed, are made difficult 
by this lack of documentation~\cite{ref:MSF1,ref:Galan}.
The desired range~$R$ of an artillery piece depends on defense capabilities and local 
topography, and the desired energy~$T$ on building methods and materials.
The actual assessment of the necessary capacity~$(R,T)$ must be made by experts 
in military history, but when this is done the dimensions of a weapon, that
is optimal according to the principles presented here, follow from only a few simple 
algebraic calculations. 
\section{Trebuchet}\label{sec:Treb}
\begin{figure}[htb]
	\centering	
	\includegraphics[width=0.95\textwidth]{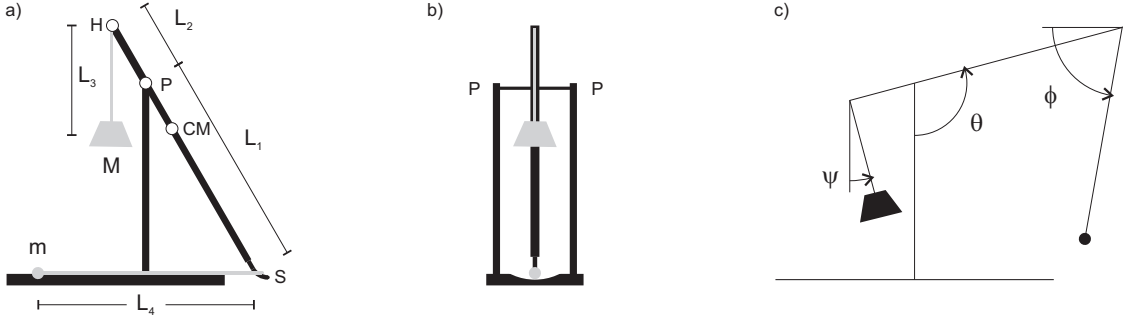}
		\caption{
		Trebuchet.
		a)~Initial configuration with projectile in trough and target to the left.
		H: hinge. P: pivot. S: spigot.
		b)~Same, but seen from target.
		c)~Snapshot of movement with generalized angular coordinates.}
		\label{fig:Trebuchet_Scaling}
\end{figure}

The basic design and functionality of a trebuchet was described recently
from different perspectives by Saimre~\cite{ref:TS} and Denny~\cite{ref:Denny}.
Here, we give only a brief account that refers to the schematic diagram 
in~\fref{fig:Trebuchet_Scaling}a where the necessary variables are
identified.
The beam, which extends from H to S, is divided into two sections by a pivoting 
axle P-P seen best in~\fref{fig:Trebuchet_Scaling}b.
The long section~$L_1$ carries a sling of length~$L_4$, and a counterweight of
mass~$M$ is hinged at H to the short section~$L_2$ by an arm of length~$L_3$.
The projectile has mass~$m$ and is placed in a pouch with two ropes that 
constitute the sling.
A spigot~S at the end of the long beam section is part of a clever release 
mechanism for the projectile.
The center of mass for the beam is at~CM and the mass is~$m_b$.
The beam is treated as a rigid and thin cylinder with diameter~$D$ and moment of 
inertia relative to P given by~$\I_b=\frac{1}{3}m_b(L_1^2-L_1L_2+L_2^2)$.
The distance from P to CM is~$L_{CM}=(L_1-L_2)/2$. 

When the trebuchet is fired, gravity pulls~$M$ down so beam and sling rotate 
about the pivot and spigot, respectively.
A snapshot of the motion is shown in~\fref{fig:Trebuchet_Scaling}c.
The projectile has been lifted off the trough and it is released into a ballistic 
path towards a target to the left at a later time determined by the spigot setting.
\Fref{fig:Trebuchet_Scaling}c also identifies three angles, which are the
generalized coordinates of the internal movement. 

The counterweight and projectile are treated as point particles.
The total rotational kinetic energy is then~$T_r=\frac{1}{2}\I_b\dot\ta^2$, 
where dot-notation is used for differention with respect to 
time,~$\dot\ta=d\ta/dt$. 
The total translational kinetic energy~$T_t$ is the sum of the kinetic energies 
of the three centers of mass.
Each has the form~$\frac{1}{2}m\dot\rvec^2$, where the velocity~$\dot\rvec$ 
depends on time through the angles~$(\ta,\psi,\phi)$ and their 
derivatives~$(\dot\ta,\dot\psi,\dot\phi)$.
The total kinetic energy of the trebuchet is~$T=T_r+T_t$, and the total potential
energy~$U$ is the sum of the potential energies of the three centers of mass.
$U$ depends on time through the angles, but is independent of angular speeds.
The kinetic and potential energies determine the equations of motion that 
govern the internal movement. 
This has three phases:
The projectile slides in the trough during phase~I, which lasts as long as the 
supporting normal force differs from zero. 
It is lifted and rotates relative to the spigot during phase~II, 
and phase~III starts when the projectile is released from the pouch.

The equations for the inner movement determine~$\ta$,~$\psi$ and~$\phi$ 
as functions of time for given initial conditions and parameters for the engine. 
These are, respectively 
\begin{eqnarray}\label{eq:InitialConditions}
	\ta_i,\quad\psi_i,\quad\phi_i,\quad\dot\ta_i,\quad\dot\psi_i,\quad\mathrm{and}\quad\dot\phi_i~,
\end{eqnarray}
and
\begin{eqnarray}\label{eq:Parameters}
	L_1,\quad L_2,\quad L_3,\quad L_4,\quad D,
	\quad M,\quad m\quad\mathrm{and}\quad m_b~.
\end{eqnarray}
The many variables imply a multitude of machines and this must be reduced.
Historical trebuchets were never propped and always at rest prior to a shot, 
so~$\psi_i=\phi_i=0$ and~$\dot\ta_i=\dot\psi_i=\dot\phi_i=0$.
The change of potential energy from a shot is initiated
and until the engine has come to rest again is
\begin{eqnarray}\label{eq:DU}
	\D U=(ML_2-m_bL_{CM})g(1+\cos\ta_i)~,
\end{eqnarray}
where~$g$ is the local gravitational acceleration. 
The initial beam angle~$\ta_i$ should be as small as practically possible to 
maximize~$\D U$.
We have chosen~$\ta_i=30^0$ as a reasonable standard.

An engine works best when~$M$ falls almost vertically for as long as possible, 
and this implies a long arm~$L_3$. 
The longest is~$L_2+L_3=H$, where~$H=L_1\cos\ti$ is the height of the pivot.
A more realistic choice is~$L_2+L_3=FH$ with~$F=2/3$ or~1/2.
Both values are examined to illustrate how much~$F<1$ impairs the performance.

Seven of the~14 parameters in~\eref{eq:InitialConditions} and~\eref{eq:Parameters}
are now fixed, and the diameter~$D$ is a function
of beam mass and length for a given density of the wood used. 
Six free parameters therefore remain.
They are
\begin{eqnarray}\label{eq:FreeAbsParameters}
	L_1,\quad L_2,\quad L_4,\quad M,\quad m\quad\mathrm{and}\quad m_b~.
\end{eqnarray}
\section{Scaling, fewer free parameters and an internal force}\label{sec:Sca}
The masses and lengths in the equations of motion can be measured such they become 
dimensionless. 
We choose to measure masses in units of the counterweight mass~$M$ 
and lengths in units of the short beam section~$L_2$.
The units of time, speed, energy and force are 
then~$(L_2/g)^{1/2}$,~$(L_2g)^{1/2}$,~$MgL_2$ and~$Mg$, respectively.
The dimensionless form of the coupled differential equations for the internal 
movement are written explicitly in~\ref{app:EOM}.
They include the five parameters
\begin{eqnarray}\label{eq:Design}
	l=\frac{L_1}{L_2},\quad
	l_4=\frac{L_4}{L_1},\quad
	L=\frac{L_3}{L_2},\quad
	\mu_b=\frac{m_b}{M}
	\quad\mathrm{and}\quad
	\mu=\frac{m}{M}
	~,
\end{eqnarray}
but~$L=Fl\cos\ti-1$ is not a free parameter, so these are the four
\begin{eqnarray}\label{eq:4free}
	l,\quad
	l_4,\quad
	\mu_b
	\quad\mathrm{and}\quad
	\mu
	~,
\end{eqnarray}
where the first three are design parameters that refer to the engine as such, 
and~$\mu$ relates to the projectile.
The efficiency of the engine~$\en$ and, for the projectile, the scaled 
range~$\rho$ and kinetic energy at target~$\tau_e$ are
\begin{eqnarray}\label{eq:Aux}
	\en=\frac{T}{\D U},
	\quad
	\rho=\frac{R}{L_2},
	\quad\mathrm{and}\quad
	\tau_e=\frac{T}{MgL_2}
	~,
\end{eqnarray}
where
\begin{eqnarray}\label{eq:T}
	T=\frac{1}{2}mv_r^2+mgh_r~.
\end{eqnarray}
Here~$v_r$ and~$h_r$  are the projectile's speed and height at release, 
respectively.
The efficiency defined in~\eref{eq:Aux} reaches one if the engine can 
be adjusted such that beam and counterweight are permanently at rest 
at the end of phase~II.
Note that efficiency is defined differently in~\cite{ref:TS,ref:Denny}.

The equations of motion can be solved only by numerical methods except in 
special cases.
The first instant of the motion is such a case.
When the initial values of angles and angular speeds are inserted 
in~\eref{eq:EOM_I_1} and~\eref{eq:EOM_I_2}, the equations become algebraic
in the angular accelerations~$\ddot\ta$ and~$\ddot\psi$ and are easily solved.
The initial acceleration of the counterweight~$\avec_{CW}$ can then be 
calculated.
It points straight down as in a free fall, but the acceleration is reduced 
relative to gravity.
The total force on the counterweight equals mass times 
acceleration and is the sum of (downward) gravity~$M\gvec$ and (upward)
reaction from the hinge~$\Fvec_H$,
\textit{i.e.}~$M\avec_{CW}=M\gvec+\Fvec_H$.
Therefore,~$\Fvec_H=-M\gvec$ before a shot is initiated~$(\avec_{CW}=\mathbf{0})$,
but immediately after
\begin{eqnarray}\nonumber
	F_H &=
		-M\gvec\left(1-
		\frac{L_2\sin^2\ti(ML_2-m_bL_{CM})}
		{ML_2^2\sin^2\ti+\I_b+mL_1^2\cos^2\ti}
		\right) \\\nonumber
		&=
		-M\gvec\left(1-
		\frac{\sin^2\ti(1-\frac{1}{2}\mu_b(l-1))}
		{\sin^2\ti+\frac{1}{3}\mu_b(l^2-l+1)+\mu l^2\cos^2\ti}
		\right)    \\\nonumber
		&=
		-M\gvec\left(1-0.42\right) 
		\quad\mathrm{with}\quad
		\ti=45^0,~l=5,~\mu_b=6\%~\mathrm{and}~\mu=1\%
		~.
\end{eqnarray}
This exemplifies the calculation of an internal force from 
the internal movement.
\section{Strength of beam and relation between lengths and masses}
\label{sec:SoB}
A good trebuchet transfers most of the counterweight's potential energy to the 
projectile.
Some is used, however, to raise and move the center of mass of the beam, so a 
light beam is desirable, but the beam must also be sufficiently strong.
In preparation for a shot, the counterweight is lifted by the short end of the 
beam as the long end is being pulled down.
The load is strongest when the beam is horizontal, and here it can be 
considered quasi-static and clamped at the pivot.
The bending of the two beam sections are then~\cite{ref:GG}
\begin{eqnarray}\label{eq:bend_L2}
	\be_1
	=\frac{MgL_1L_2}{3\M_e\I}
	\quad\quad\mathrm{and}\quad\quad
	\be_2
	=\frac{MgL_2^2}{3\M_e\I}
	\quad\quad\mathrm{with}\quad\quad
	\I=\frac{\pi}{64}D^4
	~,
\end{eqnarray}
where~$u_1=\be_1L_1$ and~$u_2=\be_2L_2$ are the deflections at the 
end points of the long and the short beam sections, respectively.
$\M_e$ is Young's modulus of elasticity for the wood being used for the beam
and~$\I$ the second moment of area for the beam profile.
Lengths and masses are independent parameters in the equations of motion, 
but~\eref{eq:bend_L2} with a given~$\be_2$ forms a bond between~$M$ 
and~$L_2$ that imply mutual scaling of masses and limear dimensions.
The beam diameter~$D$ is given by
\begin{eqnarray}\label{eq:BeamMass}
	m_b=\rho_d\frac{\pi}{4}D^2(L_1+L_2)~,
\end{eqnarray}
where~$\rho_d$ is the density of the selected wood and this leads to a 
particular scaling for~$D$.

The strain~$S$ at a position along the beam is half the diameter,~$D/2$, times the 
curvature which is largest at the pivot where
\begin{eqnarray}\label{eq:Strain}
	S=\frac{D}{2}~\frac{MgL_2}{\M_e\I}=\frac{3D}{2L_2}\be_2
	~.
\end{eqnarray}
The stress~$\sigma$ of the beam is related to the strain by~$\sigma=\M_eS$, and
the modulus of rupture is the upper limit for the stress beyond which the wood breaks.
This is often near~1\% of~$\M_e$, so to stay on the safe side we take~$S^{max}\ll1/100$.
A typical value of~$D/L_2$ is~1/3 so~$S^{max}\simeq1/600$ for~$\be_2=1/300$,
which is taken to be the maximum allowed bending and used throughout except when the value
is specified explicitly.
It is possible to use fixed~$S^{max}$ instead of fixed~$\be_2$ to control beam 
strength.
This implies a different scaling, however, and is not pursued here.

Young's modulus and the density of strong wood are taken to 
be~$\M_e=12$GPa and~$\rho_d=700$kg/m$^3$, respectively.
The maximum bending, a dimensionless constant~$K$ and two useful 
material constants that depend on gravity are
\begin{eqnarray}\label{eq:kMkD}
	\be_2&=1/300
	\hspace{15pt}
	&\hspace{30pt}K=\frac{l+1}{\mu_b}																															\nonumber\\
	k_M&=\frac{4\pi}{3}\frac{g\rho_d^2}{\M_e}&=1.68\cdot10^{-3}\mathrm{~kg/m}^4 \nonumber\\
	k_D&=\frac{16}{3}\frac{g\rho_d}{\M_e}&=3.06\cdot10^{-6}\mathrm{~1/m}
	~.
\end{eqnarray}

The bond between~$L_2$ and~$M$ in~\eref{eq:bend_L2}, the expression
for the beam mass in~\eref{eq:BeamMass} and the constants in~\eref{eq:kMkD}
lead to scaling relations for~$L_2$,~$M$,~$T$ and~$D$
\begin{eqnarray}\label{eq:ExtraScalings}
	M=K^2\frac{k_M}{\be_2}L_2^4~,
	\quad
	T=K^2\frac{k_M}{\be_2}g\tau_e L_2^5
	\quad\mathrm{and}\quad
	D=\sqrt{K\frac{k_D}{\be_2}}~L_2^{3/2}
	~.
\end{eqnarray}
The remaining parameters scale according to the relations 
in~\eref{eq:Design} and~\eref{eq:Aux}.
Thus, lengths are proportional to~$L_2$, masses to~$L_2^4$, 
energies to~$L_2^5$ and~$D$ to~$L_2^{3/2}$.

For a selected material and allowed bending,
we have now seen that given values of~$(l,l_4,\mu_b,\mu)$ and a single 
absolute parameter determine an engine completely, but a problem arises
with a given capacity~$(R,T)$, because not only one but two absolute 
parameters are given.
This leads to inconsistent values of other absolute parameters including~$m$.
The situation may be resolved iteratively, however, if the inconsistencies
are not too big:
The kinetic energy and range of the projectile are related approximately 
by~$T\simeq\frac{1}{2}mgR$, so if~$m$ is off by~$\D m$ at the given~$R$, 
then~$\D T/T\simeq\D m/m=\D\mu/\mu$,
and~$\mu$ can then be adjusted (iteratively if needed) to deliver the 
desired~$T$.
A given capacity~$(R,T)$ thus necessitates adjustment of~$\mu$ 
and reduces the number of free parameters from four to three.  
Random selection of~$(l,l_4,\mu_b)$, however, is not likely to result 
in a particularly good design, so a criterion for optimization is needed.
\section{Quality function~$\Q$, force factor~$\F$ 
				and optimization procedure}\label{sec:Q_func}
Two important, dimensionless variables have not yet been defined.
One is the quality function~$\Q$ and the other a measure of the reaction 
forces within the engine~$\F$. 
We first consider~$\Q$. 
It must increase with parameters such 
as~$\trho=R/L_1$,~$\mu\simeq2T/(MgR)$ and~$\en=T/\D U$, 
because they measure, respectively, range in terms of size~$L_1$, 
kinetic energy relative to weight~$M$ for a given range~$R$, 
and efficiency, but~$\Q$ must also be a decreasing function of internal 
forces that cause wear and tear on the engine.
It could therefore have the form
\begin{eqnarray}\label{eq:Q}
	\Q=\frac{\trho\mu\en}{\F}~,
\end{eqnarray}
where~$\F$ needs to be defined.
The balance between the factors in~\eref{eq:Q} can be changed by raising them 
to suitable powers different from one, and other factors can be included.
We continue with~\eref{eq:Q} and take it as the definition of~$\Q$.

The factor~$\F$ must represent all internal forces.
This is done by selecting the sling tension~$\Fvec_S$ and 
the force on the frame at the fulcrum~$\Fvec_R$. 
The tension $\Fvec_S$ is measured by the largest value of the component 
perpendicular to the beam~$F_{S\perp}$, which is reached shortly before
the release time~$t_r$. 
The reaction~$\Fvec_R$ is measured by its maximum value~$F_R$.
This is seen at the time~$t_m$ when the initial fall of the counterweight is 
slowed down the most, 
but we also include the horizontal component~$F_{Rh}$ of~$\Fvec_R$.
This goes through two extrema in opposite directions and in rapid 
succession.
The extrema are most often of comparable size and measured by their numeric 
sum~$\D F_{Rh}=|F_{Rh}(t_1)|+|F_{Rh}(t_2)|$, where~$t_1<t_m$ and~$t_2>t_m$. 
With these variables, $\F$ is defined by the geometric mean
\begin{eqnarray}\label{eq:F}
	\F=\left(\left(\frac{F_{S\perp}}{mg}\right)^2
	\frac{F_R}{Mg}~\frac{\D F_{Rh}}{Mg}\right)^{1/4}~,
\end{eqnarray}
which balances sling tension and reaction forces.
The quantity~$\D F_{Rh}$ measures the forces that tend to tilt and move the 
engine during a shot.
If this were allowed, it would be a sink of mechanical energy but 
is prevented by a sufficiently strong and heavy support, often in the form 
of a large trestle.

When the equations of motion are integrated, the inner movement of the 
machine is known completely.
All internal forces and torques including those in~\eref{eq:F} can then be 
determined by the use of Newton's second laws for translation and rotation, 
respectively.
One always finds~$F_{S\perp}\gg mg$, so the apparent 
weight~$m_{dyn}=F_{S\perp}/g$ carried 
by the long beam section rises to values much larger than~$m$.
$F_R$ is typically a few times~$Mg$,~$\D F_{Rh}$ always amounts to a large 
fraction of~$Mg$.

The equations of motion do not include the release mechanism for the projectile. 
The integration in phase~II can therefore be extended sufficiently in time to 
include the maximum of~$\Q$, 
and any moment can subsequently by perceived as a virtual time for release.
The instant~$t_r$ at which~$\Q$ reaches its maximum is taken as the 
actual release time, and the scaled parameters in~\eref{eq:Aux} are evaluated 
here.

Finding the maximum of~$\Q$ for given capacity~$(R,T)$ and maximum 
bending~$\be_2$ implies following a procedure that may seem complicated 
and tedious, but it is not difficult to implement numerically.
It includes the following steps
\begin{enumerate}[\quad 1)]
	\item			
			A box in~$(l,l_4,\mu_b)$-space and values~$\mu_1$ and~$\mu_2$ are 
			carefully selected to match the desired capacity~$(R,T)$ and 
			bending~$\be_2$.
			This requires exploratory calculations.
	\item 
			A design~$(l,l_4,\mu_b)$ is chosen at random within the box and
			the equations of motion solved twice. 
			First with~$\mu_1$ returning~$T_1$, 
			then~$\mu_2$ returning~$T_2$.
	\item  
			$\mu_1$ and~$\mu_2$ are selected in step~1 such that $T_1<T<T_2$ so
			linear interpolation determines an improved~$\mu=\mu_3$ 
			that returns~$T_3$.
	\item 
			This is iterated until~$|T-T_i|/T<10^{-3}$.
			Good selections in step~1 makes the convergence fast.
			$\Q$ is calculated after the last iteration.
	\item 
			A new set~$(l,l_4,\mu_b)$ is selected as in step~2 and the
			procedure repeated until the maximum of~$\Q$ is localized.
\end{enumerate}
\section{Optimization in vacuum}\label{sec:Opt}
\begin{figure}[htb]
	\centering	
	\includegraphics[width=0.65\textwidth]{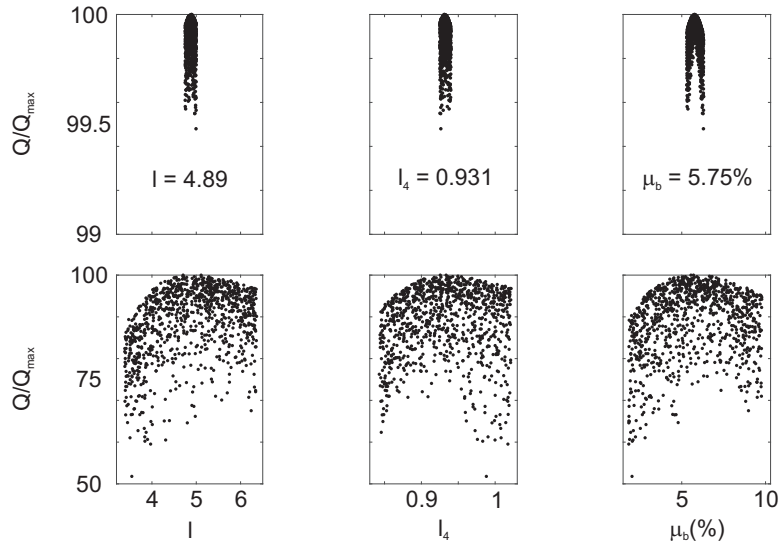}
		\caption{
		$\Q/\Q_{max}$ in~\% as function of~$l$,~$l_4$ and~$\mu_b$.
		Upper panels, narrow intervals around maximum and small drops.
		Lower panels, wider intervals and much larger drops.
		$F=1$ and capacity:~$R=240$m and~$T=26$kJ.}
		\label{fig:Merit_Ms}
\end{figure}

The procedure for localizing the maximum of~$\Q$ is illustrated
in~\fref{fig:Merit_Ms}. 
The three figures in the lower panel show the same~$10^3$ calculated~$\Q$-values 
for randomly selected design~$(l,l_4,\mu_b)$ projected on the respective planes.
The intervals are broad and relative drops down to~50\% of a clear maximum are 
seen.
The three figures in the upper panel also show~$10^3$ values of~$\Q$, but for 
parameters selected from much narrower intervals to localize the maximum more 
precisely.
The intervals show that~$\Q$ depends much more strongly on~$l_4$ than on~$l$ or~$\mu_b$, 
and a fit of the~$\Q$-values in the upper panel by a quadratic form shows that
the principal axis in the direction of maximum curvature lies very close to the
direction of~$l_4$. 

A total of~$\simeq10^4$ integrations were required to accumulate 
the~$2\times10^3$ points in the figure:                        
For each randomly selected design~$(l,l_4,\mu_b)$ it took on the average~5 
integrations to adjust~$\mu$ for the desired capacity~$(R,T)$.
These are the iterations mentioned in step~4 of the procedure 
in~\sref{sec:Q_func}.
Many more than~$10^4$ integrations were actually done because exploratory 
integrations are also required as mentioned in step~1. 

\Fref{fig:Merit_Ms} serves as an illustrative example, but a slightly different 
and more efficient practice was followed in most calculations.
The interpolations in step~3 require fewer steps when~$T$ is replaced 
by~$\log(T)$, and relatively wide parameter intervals at the 
outset were narrowed systematically around a running estimation of 
the maximum position such that optimal parameter values could be found with 
sufficiently accurately already after a thousand integrations or so.
\subsection{Analytical expressions for absolute parameters}\label{sec:AEAP}
\begin{figure}[htb]
	\centering	
	\includegraphics[width=0.4\textwidth]{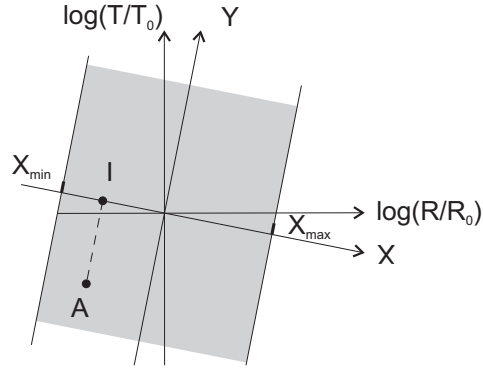}
		\caption{Coordinates~$(X,Y)$.
		Scaling along lines of constant~$X$.}
		\label{fig:XY}
\end{figure}

\begin{figure}[htb]
	\centering	
	\includegraphics[width=1\textwidth]{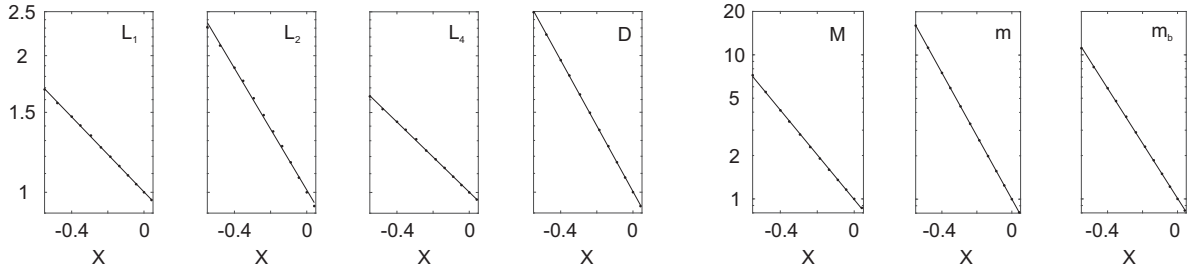}
		\caption{
		Full points, calculated values of ~$(Z/Z_0)/(R/R_0)^{n_Z}$~\textit{vs}~$X$.
		Straight lines, fits by~$f_Z(X)$.
		Semi-logarithmic plots.
		$R_0=240$m.
		$T_0=26$kJ.
		$\ti=30^0$.
		$F=1$.}
		\label{fig:fZ_vs_X}
\end{figure}

According to the results on scaling in~\sref{sec:Sca}, the optimal design 
at~$(R_0,T_0)$ is also the best along the curve~$T/T_0=(R/R_0)^5$ in 
the~$(R,T)$-plane.
This is a straight line in a logarithmic representation and a rotation
expressed in matrix form by
\begin{eqnarray}\label{eq:XY_RT}	
	\left\{
		\begin{array}{c}
			X \\
			Y
		\end{array}
	\right\}
	&=&\frac{1}{\sqrt{26}}
	\left\{
		\begin{array}{cc}
			5 & -1 \\
			1 & 5
		\end{array}
	\right\}	
	\left\{
		\begin{array}{c}
			\log(R/R_0) \\
			\log(T/T_0)
		\end{array}
	\right\}
\end{eqnarray}
defines coordinates~$(X,Y)$ for which the scaling applies at constant~$X$
as illustrated in~\fref{fig:XY}. 
A finite number of optimizations along the~$X$-axis followed by 
interpolation and scaling determine optimal engines over 
a continuous region limited only by the range of $X$-values covered 
(the gray area in \fref{fig:XY}).
Suppose, as an example, that one wishes to find the best engine at the
point~A with coordinates~$(R,T)$.
Scaling connects~A to the point I with~$(X,Y)$ coordinates
\begin{eqnarray}
		X_i=
		\frac{1}{\sqrt{26}}\left(5\log\frac{R}{R_0}-\log\frac{T}{T_0}\right)
		\quad \mathrm{and} \quad
		Y_i=0~.  
		\label{eq:Xi}
\end{eqnarray}
This point will most likely not belong to the set of calculated points, but 
linear interpolation with the two neighboring points on the~$X$-axis gives 
a good estimate.
Interpolation is tedious and requires tables for each parameter. 
It can be avoided, however, by the introduction of a set of analytical 
functions~$f_Z(X)$, where index~$Z$ refers to any one of the absolute 
parameters for the trebuchet. 
We take~$M$ as an example and find because masses are proportional to 
lengths for constant~$X$~\eref{eq:ExtraScalings}
\begin{eqnarray}\label{eq:MR}
	\frac{M}{M_0}=f_M(X)\left(\frac{R}{R_0}\right)^4~,
\end{eqnarray}
where~$f_M(X)$ is found from the integrations along the~$X$-axis.
For any~$(R,T)$, the interpolation point~$X_i$ on the~$X$-axis is given 
by~\eref{eq:Xi}, so~$M$ is a function of~$(R,T)$ alone. 
It turns out that~$f_M(X)$ can be represented closely over broad 
ranges of~$X$ by the form
\begin{eqnarray}\label{eq:f_MX}
	f_M=a_M\exp\left(b_M\sqrt{26}X\right)	
\end{eqnarray}
where~$a_M$ and~$b_M$ are adjustable constants.
After insertion of~\eref{eq:f_MX} in~\eref{eq:MR} and use of~\eref{eq:Xi} 
one finds
\begin{eqnarray}\label{eq:MRT}
	\frac{M}{M_0}=a_M\left(\frac{R}{R_0}\right)^{4+5b_M}
	\left(\frac{T}{T_0}\right)^{-b_M}	
	\hspace{5mm}\mathrm{with}\hspace{5mm}
	a_M\simeq1~.
\end{eqnarray}
Expressions similar to~\eref{eq:MR},~\eref{eq:f_MX} and~\eref{eq:MRT} apply for 
all other absolute parameters~$Z$, so
\begin{eqnarray}\label{eq:f_ZX}
	\frac{Z}{Z_0}=f_Z(X)\left(\frac{R}{R_0}\right)^{n_Z}
	\quad\mathrm{where}\quad
	\left\{
	\begin{array}{ll}
			n_Z=1 	&\mathrm{for}~L_1,~L_2~\mathrm{and}~L_4	\\
			n_Z=3/2	&\mathrm{for}~D	\\
			n_Z=4		&\mathrm{for}~M,~m~\mathrm{and}~m_b	\\
	\end{array}
	\right.
\end{eqnarray}
or more explicitly
\begin{eqnarray}\label{eq:ZRT}
	\frac{Z}{Z_0}=
	a_Z\left(\frac{R}{R_0}\right)^{p_Z}
	\left(\frac{T}{T_0}\right)^{q_Z}
	\quad\mathrm{where}\quad
	\left\{
	\begin{array}{l}
			p_Z=n_Z+5b_Z \\
			q_Z=-b_Z~.
	\end{array}
	\right.
\end{eqnarray}
The result of fitting the form~\eref{eq:f_ZX} to the seven parameters are shown 
in~\fref{fig:fZ_vs_X}.
The functions~$f_Z$ are straight lines in the chosen semi-logarithmic 
representation.
They are seen to represent the calculated values accurately so the analytical 
expression~\eref{eq:ZRT} gives a very good estimate of the optimal value for the chosen 
parameter for any desired capacity~$(R,T)$ with~$X$-value within the interval covered.
All data for lengths are plotted at the same scale and so are the data for masses. 
This facilitates comparison within each group, but note that the length scale covers a 
factor~$\simeq$2.5 variation whereas the mass scale covers a factor~$\simeq$20 variation.

\Eref{eq:ZRT} is linear in a logarithmic representation.
The relation between~$(R,T)$ and two absolute parameters~$Z_i$ and~$Z_j$  
from~\eref{eq:FreeAbsParameters} can therefore be expressed by
\begin{eqnarray}\label{eq:RT_ZiZj}
	\left\{
		\begin{array}{c}
			\log(Z_i/Z_{i0})) \\\
			\log(Z_j/Z_{j0}))
		\end{array}
	\right\}
	=
	\left\{
		\begin{array}{cc}
			p_{Z_i} & q_{Z_i} \\
			p_{Z_j} & q_{Z_j}
		\end{array}
	\right\}
	\left\{
		\begin{array}{c}
			\log(R/R_0) \\
			\log(T/T_0) 
		\end{array}
	\right\}~,
\end{eqnarray} 
where we have use~$\log(a_Z)\simeq0$.
The equation can be inverted provided the
determinant~$\Dc_{ij}=p_{Z_i}q_{Z_j}-q_{Z_i}p_{Z_j}$ is different from zero,
and if so
\begin{eqnarray}\label{eq:ZiZj_RT}	
	\left\{
		\begin{array}{c}
			\log(R/R_0) \\
			\log(T/T_0)
		\end{array}
	\right\}=\frac{1}{\Dc_{ij}}
	\left\{
		\begin{array}{cc}
			q_{Z_j} 	& -q_{Z_i} \\
			-p_{Z_j} 	& p_{Z_i} 
		\end{array}
	\right\}	
	\left\{
		\begin{array}{c}
			\log(Z_i/Z_{i0})) \\
			\log(Z_j/Z_{j0})) 
		\end{array}
	\right\}~.
\end{eqnarray} 
Absolute parameters~$(Z_i,Z_j)$ and~$(Z_k,Z_l)$ are related two-by-two if~$(R,T)$ is eliminated
from~\eref{eq:RT_ZiZj} and~\eref{eq:ZiZj_RT}, and~$(X,Y)$ can be included by the use of~\eref{eq:XY_RT}.

\Tref{tab:CPNC} shows constants derived from of the fits in~\fref{fig:fZ_vs_X}. 
The interval of~$X$-values for which the constants apply is also given. 
Constants from other fits of similar quality to data with~$X>0$,~$F=2/3$ and~$\ti=40^0$ are also included.
The data with~$X>0$ and~$F=1$ show a narrow region of~$X$ from~0.3 to~0.4 with competing optimized design.
\begin{table}[tbh]\footnotesize
	\centering
	\begin{tabular}{c|cc|c|ccccccc}
			$X$		&$\ti$	&F&				&$L_1$		&$L_2$ 		&$L_4$		&$D$			&$M$		&$m$			&$m_b$ 			\\\hline
						&	& &$Z_0$ 	& 3.90m		& 0.80m		&	3.62m		& 0.22m		&2158kg	& 20.6kg	& 124kg						\\\hline
						&$30^0$	&1	&$a_Z$	& 1.00  	& 1.02   	&	1.00  	& 1.00  	&0.98   & 1.00  	& 1.01  	\\
			-0.55	&$30^0$	&1	&$p_Z$ 	& 0.0654	&-0.5013	&	0.1243  &-0.1299  &0.4830 &-0.9481  &-0.3045 	\\
			0			&$30^0$	&1	&$q_Z$	& 0.1869  & 0.3003  & 0.1752  & 0.3260	&0.7034	& 0.9896 	& 0.8609	\\\hline
			lb		&$30^0$	&1	&$a_Z$	& 0.9967	& 1.0104 	& 0.9937  & 1.0025	&0.9893	& 0.9988  & 1.0023 	\\
			0			&$30^0$	&1	&$p_Z$	& 0.0816	&-0.7853	& 0.1979	&-0.2069	&0.7431 &-0.9613  &-0.4591	\\
			0.40	&$30^0$	&1	&$q_Z$	& 0.1837	& 0.3571	& 0.1604	& 0.3414	&0.6514	& 0.9923	& 0.8918 	\\\hline
			ub		&$30^0$	&1	&$a_Z$	& 1.12 		& 1.21 		&	1.31 		& 1.05		&0.841 	& 1.01		& 1.25 		\\
			0.30	&$30^0$	&1	&$p_Z$	&-0.1385	&-1.0423 	&-0.1625	&-0.2839	&0.9490	&-0.9757 	&-0.8238 	\\
			0.57	&$30^0$	&1	&$q_Z$	& 0.2277 	& 0.4084 	&	0.2325 	& 0.35678	&0.6102	& 0.9951	& 0.9647 	\\\hline
						&$30^0$	&2/3&$Z_0$ 	& 3.97m		& 0.67m		& 3.58m		& 0.21m 	&2590kg	& 20.7kg  & 112kg		\\\hline
						&$30^0$	&2/3&$a_Z$	& 1.00  	& 1.01		& 1.00  	& 1.00		& 0.99	& 1.00		&	1.00		\\
			-0.55	&$30^0$	&2/3&$p_Z$	& 0.08611	&-0.3131  & 0.1640 	& -0.0856	&0.2839	& -0.9459 & -0.1487	\\
			0			&$30^0$	&2/3&$q_Z$	& 0.1828  & 0.2626  & 0.1672 	& 0.3171  &0.7432	& 0.9892 	& 0.8297	\\\hline
						&$30^0$	&2/3&$a_Z$	& 1.00  	& 1.00		& 1.00  	& 1.00		& 1.00	& 1.00		&	1.00		\\
			0			&$30^0$	&2/3&$p_Z$	& 0.0400 	&-0.6865 	& 0.1465  & -0.1851 & 0.6328& -0.9672	& -0.4210	\\
			0.57	&$30^0$	&2/3&$q_Z$	& 0.1920 	& 0.3373 	& 0.1707 	& 0.3371  & 0.6735& 0.9934 	& 0.8842	\\\hline
						&$40^0$	&2/3&$Z_0$ 	& 4.13m   & 0.63m   & 3.47m  	& 0.21m   & 2925kg& 20.7kg  & 116kg		\\\hline
						&$40^0$	&2/3&$a_Z$	& 1.00  	& 1.00		& 1.00  	& 1.00		& 1.00	& 1.00		&	1.00		\\
			-0.52	&$40^0$	&2/3&$p_Z$	& 0.0857 	&-0.3429  & 0.2012  &-0.0944  & 0.3080&-0.9481  & -0.1641	\\
			0			&$40^0$	&2/3&$q_Z$	& 0.1829  & 0.2686  & 0.1598  & 0.3189  & 0.7384& 0.9896  & 0.8328 	\\\hline
						&$40^0$	&2/3&$a_Z$	& 1.00  	& 1.01		& 1.00  	& 1.00		& 0.99	& 1.00		&	1.01		\\
			0			&$40^0$	&2/3&$p_Z$	& 0.0672  &-0.6765 	& 0.2546 	&-0.1858 	& 0.6100&-0.9666 	&-0.3880	\\
			0.52	&$40^0$	&2/3&$q_Z$	& 0.1866 	& 0.3353  & 0.1491  & 0.3372  & 0.6780& 0.9933 	& 0.8776
		\end{tabular}
	\caption{Constants in~\Eref{eq:ZRT} with~$R_0=240$m and~$T_0=26$kJ.
						Limits on~$X$.
						Overlap for~$0.3<X<0.4.$
						lb: lower branch.
						ub: upper branch.}
	\label{tab:CPNC}
\end{table}
\subsection{Two branches of optimized design}\label{sec:2B}
Two different optimized design compete at~$X$ between~0.3 and~0.4 when~$\ti=30^0$ 
and~$F=1$ as seen in~\tref{tab:CPNC}.
In this narrow range of~$X$, the quality function~$\Q=\Q_X(l,l_4,\mu_b)$ shows two 
maxima, which are seen most easily when the calculations are 
projected on the~$(l_4,\Q)$-plane 
(as in middle figure in lower panel of~\fref{fig:Merit_Ms}).
The maximum belonging to the lower branch with~$X_{max}=0.40$ dissolves 
near~$X_{max}$.
The other maximum on the upper branch with~$X_{min}=0.30$ emerges near~$X_{min}$,
and the two maxima are equally high at~$X=X_d$, where~$X_d\simeq0.34$.
$\Q$ is continuous at~$X_d$, but all other parameters are 
discontinuous at this point with steps that depend on~$X$ but not on~$Y$.
The size of the step~$S_Z$ follows from~\eref{eq:XY_RT} after inversion 
and~\eref{eq:RT_ZiZj}. 
It reads
\begin{eqnarray}\label{eq:ZiZj_vs_X}
	S_Z
	=
	\frac{Z_{ub}}{Z_{lb}}
	=
	\frac{a_{Z_{ub}}}{a_{Z_{lb}}}
	\exp\left(\frac{5p_{Z_{ub}}-q_{Z_{ub}}-5p_{Z_{lb}}+q_{Z_{lb}}}
		{\sqrt{26}}X_d\right)		
	~,
\end{eqnarray} 
where the constants are taken at the two branches.
The largest step is~$S_{L_4}=+16.3\%$. 
Others are~$S_{L_1}=+4.4\%$ and~$S_M=-8.7\%$ while~$m$ increases by less than~1\%.

For~$X$ near $X_d$ we find after expansion of the exponential function
\begin{eqnarray}\nonumber
	l_4
	=
	\frac{L_4}{L_1}
	=
	\left\{
		\begin{array}{cl}
		0.965\left(1+0.119(X-X_d)\right) & \mathrm{on~lower~branch} \\
		1.075\left(1-0.024(X-X_d)\right) & \mathrm{on~upper~branch,} 
		\end{array}
	\right.
\end{eqnarray} 
so the relative sling length is less than one and increasing 
on the lower branch, and larger than one and decreasing at a smaller rate 
on the upper.

The configuration of an optimized trebuchet at the best time for release 
of the projectile is also qualitatively different on the two branches.
On the lower branch, the beam has not yet crossed the vertical direction,
but on the upper branch, it has just crossed it
for the first time.
Competing designs are found in a very narrow interval of~$X$ 
near~0.58 when~$\ti=30^0$ and~$F=2/3$, and for other~$\ti$ and~$F$ as well.
\subsection{Frequency of optimized design}\label{sec:FoOD}
\Eref{eq:RT_ZiZj} can be expanded to cover the six absolute parameters~$Z_i$ 
listed in~\eref{eq:FreeAbsParameters}
\begin{eqnarray}\label{eq:RT_Z1Z2}
	\left\{
		\begin{array}{c}
			\log(Z_i/Z_{i0}))
		\end{array}
	\right\}_{6\times1}=
	\left\{
		\begin{array}{cc}
			p_{Z_i} & q_{Z_i}
		\end{array}
	\right\}_{6\times2}	
	\left\{
		\begin{array}{c}
			\log(R/R_0) \\
			\log(T/T_0) 
		\end{array}
	\right\}_{2\times1}~,
\end{eqnarray} 
where the dimensions of the matrices are given explicitly.
Thus, optimized absolute parameters are limited to a two dimensional hyperplane 
imbedded in a space of six dimensions.
This is the vector space spanned by the two column vectors in the~$6\times2$ matrix.
Good designs are found near every optimized design, so by including these the plane
expands in all directions to form an object of final but very small volume in the huge 
six dimensional space. 
This illustrates the rarity of good designs and how difficult it is to find them.

We further illustrate the difficulty by examining a standard set of four parameters 
chosen by Denny to represent a large trebuchet~\cite{ref:Denny}.
The standard parameters are
\begin{eqnarray}\nonumber
	L_b=12\mathrm{m},\quad M=10\hspace{2pt}000\mathrm{kg},\quad m=100\mathrm{kg}
	\quad\mathrm{and}~\quad 
	m_b=2000\mathrm{kg}~,
\end{eqnarray}
where~$L_b=L_1+L_2$ is the beam length.
They can be paired two-by-two in six ways and the 
capacity~$(R,T)$ was calculated for each by the use of~\eref{eq:ZiZj_RT} 
and~\tref{tab:CPNC} with~$\ti=30^0$ and~$F=2/3$.
Only one of the six pairs results in a meaningful capacity 
and a value of~$X$ within the limits in~\tref{tab:CPNC}.
This is~$M=10\hspace{2pt}000$kg and~$m=100$kg with~$R=288$m 
and~$T=152$kJ that implies~$X=-0.17$.
The optimized beam length for this capacity calculated by the use 
of~\eref{eq:ZRT} is~$\simeq6.6$m which is quite small 
when compared to the standard of~12m, and the weight of the 
beam~$\simeq475$kg is even smaller in relative terms compared to~2000kg.
The efficiency of the optimized engine is~$\simeq$92\%.

A trebuchet with the standard parameters does not maximize~$\Q$, but it
works and Denny~\cite{ref:Denny}
finds a maximum range in air of~360m and a projectile velocity at 
release of~63m/s.
This result is listed in the first row of~\tref{tab:Denny}
\begin{table}[tbh]\footnotesize
	\centering
		\begin{tabular}{c|ccc|cc|ccc|ccc|cc}
								&$\ti$		& $\be_2$	& $F$	& $\Q$	&$\en$		&$L_b$	&$H$		&$L_4$	&$M$		&$m$			&$m_b$ 		& $R$ 		& $T$			\\
								&					& \%				& 		& \%		& \%		& m 		& m			&  m		& ton		& kg 			& kg			& m 			& kJ			\\\hline
			Denny			&$45^0$		& 					& 		& 			& 			&\B{12} & 6.15	& 7.9		&\B{10}	&\B{100}	&\B{2000}	& 360			&					\\
			Present		&$45^0$		& 0.66 			&0.93 & 1.7  & 48.4 	&\B{12} & 6.15	& 7.9		&\B{10}	&\B{100}	&\B{2000}	& 425			&	224 		\\\hline
			Max $\Q$	&$45^0$		& 0.66 			&0.93 & 7.0  & 91.4 	& 8.67  & 5.25	& 6.19	&13.3 	&101 			& 629 		&\B{425} 	&\B{224}	\\
			Max $\Q$	&$30^0$		& 0.33 			& 1 	& 7.8  & 93.3 	& 7.15  & 5.23	& 5.69	&13.3 	&101 			& 647 		&\B{425} 	&\B{224}	\\
			Max $\Q$	&$30^0$		& 0.33 			&2/3 	& 6.7  & 92.2 	& 7.06  & 5.30	& 5.57	&15.7 	&101 			& 591 		&\B{425} 	&\B{224}	\\\hline
			Model 		&$30^0$		& 0.33 			&2/3 	& 6.7	 & 92.2		& 1.33  & 1.00	& 1.05	&19.9kg	&0.127 		& 0.745 	& 80.0 		& 53.1
		\end{tabular}
	\caption{Comparison of design.
	Denny: $R$ from~\cite{ref:Denny} for standard parameters.
	Present: Present results with Denny's parameters.
	Max $\Q$: Design at maximum~$\Q$ by	\textit{ab initio} calculations 
	and given~$(R,T)$.
	Model: Scaled values,~\Eref{eq:ExtraScalings}.} 
	\label{tab:Denny}
\end{table}
and data from
present \textit{ab initio} calculations are given in the second row.
The values of~$\be_2$ and~$F$ follow from Denny's data.
The range in vacuum of~425m results from a release velocity (64m/s) 
in full agreement with Denny's.
The different ranges are explained by aerodynamic losses and different 
definitions of range.
Air drag reduces the vacuum range to~380m by the same assumptions 
as used by Denny who also measures range by the distance from the release point 
to the point where the projectile dips below the release height.
The release is at a height of~20.2m and at a position~6.0m behind the pivot.
The range by Denny should therefore be compared with~380m$-$20m+6m=366m, which 
again shows good agreement.
The vacuum value of kinetic energy at target is~224kJ and the
efficiency~48.4\%.

The remaining results in~\tref{tab:Denny} show design that maximize~$\Q$ 
at the capacity~425m and~224kJ that follows from the standard parameters.
The first of these are for Denny's values of~$\ti$,~$\be_2$ and~$F$, and
it shows a shorter and lighter beam~$(L_b,m_b)$, a smaller height of the pivot~$H$ 
and a heavier counterweight~$M$.
The efficiency~$\en$ of the engine is increased by more than a factor of two, 
so loading requires less than half the energy needed for
loading the standard machine.
The present standard values of~$\ti$,~$\be_2$ and~$F$, and the more 
realistic~$F=2/3$ are used in the next two rows.
The steeper initial beam position and selection of a stronger beam results in 
further shortening of the beam, almost unchanged pivot height, still heavier 
counterweights and slightly increased efficiencies.
The efficiencies of the optimized engines in~\tref{tab:Denny} are all 
somewhat larger than~90\%.
This is in clear contrast to an apparent consensus in the literature on an 
upper limit near~70\%~\cite{ref:TS,ref:Denny}.

The last row in~\tref{tab:Denny} shows dimensions of a small trebuchet found
by scaling the one just above.
The small engine can be handled safely by only a couple of people, 
but the small mass of the projectile increases the aerodynamic sensitivity.
Approximately equal relative losses can be obtained, however, by selecting a
projectile material of higher density than stone.
A simple estimate gives~$\rho/\rho_{stone}\simeq2.3$, so iron would do.
\section{Trebuchets of archaeological interest}\label{sec:RoAI}
The cases selected to illustrate the present results 
were chosen by their archaeological significance.
The first relate to the sieges of Alcalá la Vieja in 1118
by Christian forces and of Saone in 1188 by Saladin's army.
Others are inspired by drawings in the sketchbook of Villard de 
Honnecourt~\cite{ref:VdH} from the early~13th~century.

The optimizations most often take outset in desired ranges~$R$ and known
or assumed projectile masses~$m$.  
Optimized parameters can then be found by the use of~\eref{eq:ZRT} to first 
determine~$T$, whereafter the remaining parameters follow 
from~\eref{eq:RT_ZiZj}.
Another starting point is an assumed beam length and partitioning, which
determines~$L_1$ and~$L_2$.
The capacity then follows from~\eref{eq:ZiZj_RT} and the remaining optimized 
parameters from~\eref{eq:RT_ZiZj}.

The results show strong variation of mass with 
size,~\textit{i.e.}~$M\propto L_2^4$ by~\eref{eq:ExtraScalings}.
This is not immediately understandable, so the optimized design to follow 
are most likely smaller and heavier than expected.
Simple proportionality,~\textit{i.e.}~$M\propto L_2$, is more straightforward, but is not 
consistent with the present analysis.
See chapter 2 of~\cite{ref:MSF2}
for a recent discussion of scaling based on intuition and~\textit{ad hoc} estimates.
\subsection{Alcalá la Vieja and Saone}\label{sec:AVS}
The ranges listed in the first three rows of~\tref{tab:Alcala_la_Vieja} are central to the discussion 
by Galán  and Galán~\cite{ref:Galan} of the siege of Alcalá la Vieja, where strategically 
advantageous locations for trebuchets were examined outside the range of arrows 
fired from the fortress~(\textit{i.e.}~$\simeq$140m).
\begin{table}[tbh]\footnotesize
	\centering
		\begin{tabular}{cccc|cccccc|cccc}
			Site	&Level&R	& T 	&$L_1$	&$L_2$	&$L_4$&$D$		&$M$	&$m_b$& $\en$	&$v$	&$h$	&$A$ 			\\
						&m		&m	& kJ	&m			& m 		& m		&  m		&tons	&kg		& \%		&m/s	& m 	&					\\\hline
			Ca		&(-15)	&\B{150}&40.3	&4.13 	&0.88 	&3.55	&0.25 	&3.11	&174 	& 90.0	&37.7 &10.0 &40.3$^0$	\\
			Ma		&(-54)	&\B{480}&124 	&5.50 	&0.71 	&5.15	&0.31 	&11.5 &334 	& 92.8	&68.5 &13.8 &39.1$^0$ \\
			Ma		&-54		&\B{426}&111 	&5.36 	&0.75 	&4.96	&0.31 	&9.70 &320 	& 92.5	&64.5 &13.5	&39.1$^0$ \\\hline
			Sa		&(-20)	&\B{250}&66.0	&4.72 	&0.85 	&4.19	&0.28 	&5.16	&242 	& 91.4	&49.1 &11.7 &39.7$^0$
		\end{tabular}
	\caption{Optimized trebuchets with~$\ti=30^0$,~$F=2/3$ and~$m=\B{50}$kg stones.
	First three rows: Siege of Alcalá la Vieja with
	Catalanes site (Ca), and Malvecino site (Ma).
	Fourth row: Siege of Saone.
	Level: Level of target relative to trebuchet.}
	\label{tab:Alcala_la_Vieja}
\end{table}
Stones found in the area and most likely used as projectiles typically weigh 
around~50kg.
The energies~$T$ that follow from~$R$ and~$m$ give values of~$X$ within the limits 
in~\tref{tab:CPNC}.
Algebraically determined optimized parameters then found are given in the 
central columns of~\tref{tab:Alcala_la_Vieja}.
The parameters to the right are efficiencies~$\en$ and initial conditions for
the projectile's ballistic trajectory towards the target, which are speed~$v$, height~$h$
and climb~$A$ at release. 
The engines are elevated relative to the targets, but this is ignored in most cases 
(and clarified by placing the target level in parenthesis).
The lower level of the targets then more than compensates for aerodynamic losses of range and 
energy along the projectile path.
In the example without parenthesis, the aim is shortened by the difference of level, and
optimized parameters are then found as before by assuming no difference of level. 
The projectile flies past the shortened aim of~426m and reaches the lower level of the 
target at~481m. 
The approximate procedure thus overshoots the range by just one meter.
The kinetic energy increases from~111kJ at zero level to~137kJ at target due to gravity.
The potential energy stored in the engine before the shot is~111kJ/0.925=120kJ, 
\textit{i.e.} less than the energy at target.

The trebuchets erected at the Catalanes site were hybrids of 
traction and fixed counterweight engines with beam lengths 
near~9.5m~\cite{ref:Galan}.
The optimized realization in~\tref{tab:Alcala_la_Vieja} is much smaller. 
Having had a small engine like this at their disposal would have been a great advantage 
for the leaders of the campaign because it makes transportation and erection much easier.
The heavy counterweight of~$\simeq3$~metric tons can be a light box that is 
taken apart for transportation, then reassembled and loaded with rocks and soil 
at a new deployment.

The Melvecino site was ruled out~\cite{ref:Galan} because of the long distance to
the fortress, but a trebuchet of manageable size placed at Malvecino could indeed have 
bombarded the fortress with 50kg stones.
However, the required counterweight of~11.5~tons (or~9.7~tons) is challenging.
The siege of Alcalá la Vieja took place early in the 12th century when counterweight 
trebuchets were still under development and perhaps had not yet matured into a stage
where such heavy counterweights could be used.

The data in the fourth row of~\tref{tab:Alcala_la_Vieja} relate to the siege of
Saone.
A likely position for trebuchets exists at a distance of~250m from the fortifications 
of the castle~\cite{ref:MSF1}, and an optimized engine that covers this range is 
seen to deliver~66kJ to the target.
In a discussion of the siege by Fulton~\textit{et al.}~\cite{ref:MSF1} it is concluded 
that the trebuchets deployed there could not have been breaching weapons at this early 
stage of development, but we have seen that well constructed engines of limited size 
and weight could indeed have bombarded the fortifications by 50kg stones arriving 
at~$\simeq185$km/h.

The siege of Saone took place only~70 years after the siege of Alcalá la Vieja, 
but they are not easily compared.
The fall of Alcalá la Vieja, located close to Madrid in Spain, was part of the 
reconquest of Spain whereas Saone, located in Syria close to the Mediterranean Sea, 
was taken during Saladin's Syrian campaign.
\subsection{Sketchbook of Villard de Honnecourt}\label{sec:SVdH}
The sketchbook of Villard de Honnecourt, which dates back to the early 13th century, 
has a drawing of the sole (or base framing) of a trebuchet and a description of a 
chest full of earth which is the counterweight. 
Both have detailed length indications.
The sizes of sole and chest suggest beam lengths and counterweights of up to~10m
and thirty metric tons, respectively.

Chevedden suggests that a huge trebuchet like this can throw	a 100kg stone 
more than~400m~\cite{ref:Chev}.
Following this suggestion, we take~$m=100$kg and $R=450$m.
The energies~$T$ that follow allow the optimized parameters in the first rows 
of~\tref{tab:SketchBook-I} to be found algebraically.
\begin{table}[tbh]\footnotesize                                                            
	\centering
		\begin{tabular}{c|cccc|ccccc|ccccc} 
			$n$	&$F$ 	&$\ti$	& $R$ 		& $T$ 		&$L_1$ 		&$L_4$		& $M$ 	& $m$			&$m_b$	& $L_b$ 	&$l$			& $\en$		& $\Q_n$	\\
					&			&     	& m   		& kJ 			& m 			&  m			&tons		&kg				& kg		&m  			& 				& \%			& \% 		\\\hline
			1		&1 		&$30^0$	&\B{450}	& 235 		& 6.13		& 5.80 		& 14.3	&\B{100}	& 660 	& 7.21		& 5.68		&	93.1 		&	7.41	\\
			1		&2/3	&$30^0$	&\B{450}	& 234 		& 6.20		& 5.69 		& 16.9	&\B{100}	& 601 	& 7.12		& 6.77		&	92.6		&	6.43	\\
			1		&2/3	&$40^0$	&\B{450}	& 234 		& 6.48		& 5.55 		& 19.1	&\B{100}	& 622 	& 7.35		& 7.51		&	91.6		&	5.55	\\\hline
			5		&1		&$30^0$	&\B{450}	&\B{235}	& 6.31		& 6.68 		& 15.6	& 101			& 611 	& 7.26		& 6.64		&	97.5		&	6.38	\\
			0		&1		&$30^0$	&\B{450}	&\B{235}	& 6.60		& 6.04 		& 13.4	& 98.3		& 807 	& 7.85		& 5.29		&	87.8		&	8.11
		\end{tabular}
	\caption{	Design relating to Villard de Honnecourt.
						Quality function~$\Q_n$ in~\eref{eq:Q_n}.}
	\label{tab:SketchBook-I}
\end{table}
$F=1$ in the first row is the ideal case and $F=2/3$ in the next two the more practical.
The initial beam angle~$\ti=40^0$ is perhaps also more appropriate then~$\ti=30^0$.
The beam lengths close to 7m seem short when compared to the sole, and 
counterweights of less than~20 tons are also relatively small in comparison with the maximum 
of~30 tons.
The variations of linear size in going from $F=1$ to 2/3 and from $\ti=30^0$ to $40^0$ are modest,
but~$M$ increases more significantly.
Finally, the counterweight was propped in row 2. 
This has only a small effect on~$\en$ with a maximum of~93.0\% at~$\psi_i=-4.6^0$
because the initial counterweight motion is slightly better.

The two lowest rows in~\tref{tab:SketchBook-I} show optimized design obtained 
by the use of the modified quality function
\begin{eqnarray}\label{eq:Q_n}
	\Q_n=\frac{\trho\mu\en^n}{\F}
	\quad\quad\mathrm{with}\quad\quad
	n\geq0~,
\end{eqnarray}
which is the same as~\eref{eq:Q} when~$n=1$.
There is more emphasis on efficiency~$\en$ when~$n>1$ and less 
when~$n<1$.
We see that~$\en$ increases from~93.2\% at~$n=1$ to~97.5\% at~$n=5$ 
and decreases to~87.8\% at~$n=0$ with only small adjustments of the engine.
The most remarkable is the change of relative sling length~$l_4$ to a value 
larger than one at~$n=5$.
This reveals a transition to a qualitatively 
different engine
(see discussions in~\sref{sec:AEAP} and~\ref{app:ME}).

Many trebuchets built early on could throw heavy stones a good distance, but 
their precise dimensions are unknown.
A trebuchet with proportions derived from the dimensions of the sole in the
sketchbook is shown in a semi-realistic drawing included as figure 26 in a commented 
facsimile version of the sketchbook published in~1859~\cite{ref:VdH}.
As measured on the drawing, the beam is divided by the pivot into two lengths 
with ratio~$l=L_1/L_2\simeq4.5$, and the initial beam angel is~$\ti\simeq25^0$, 
but figure~27 in~\cite{ref:VdH}, which serves to illustrate an explanation of the action 
of a trebuchet, shows a different initial configuration with~$\ti\simeq40^0$.
\Tref{tab:SketchBook-II} shows optimized design that comply with these findings.
\begin{table}[tbh]\footnotesize
	\centering
		\begin{tabular}{c|cccc|ccccc|ccccc} 
			\#&$F$ 	&$\ti$	& $R$ 		& $T$ 		&$L_1$ 		&$L_4$		& $M$ 	& $m$			&$m_b$	& $L_b$ 	&$l$			& $\en$		& $\Q$	\\
			 &    	&     	& m   		& kJ 			& m 			&  m			&tons		&kg				& kg		&m  			& 				& \%			& \% 		\\\hline
			1&1   	&$25^0$	&	426			& 677			& 7.36		& 7.00    & 26.8 	& 301   	& 1696	&\B{9}  	&\B{4.5}	& 93.3  	& 10.4	\\					
			2&1     &$30^0$	& 407 		& 641  		& 7.36		& 6.80		&	26.0	& 298 		&	1682	&\B{9}  	&\B{4.5}	&	92.9  	&	10.2	\\
			3&1    	&$40^0$	& 341			& 540			& 7.36		& 6.20 		& 23.7	& 299			& 1589	&\B{9}  	&\B{4.5}	&	91.5		& 10.2	\\\hline 
			4&1			&$30^0$ &	540			& 700			& 7.62		&	7.16		& 33.4	&	248			& 1605	&\B{9}  	&\B{5.5}	&	93.2		&	7.71	\\
			5&2/3		&$30^0$ &	401			& 726			& 7.62		&	6.76		& 34.9	&	343			& 1653	&\B{9}  	&\B{5.5}	&	91.6		&	8.79	\\\hline 
			6&2/3		&$30^0$ &	482			& 752			& 7.74		&	6.98		& 38.7	&	299			& 1623	&\B{9}  	&\B{6.0}	&	92.0		&	7.55	\\
			7&1/2		&$30^0$ &	297			& 858			& 7.69		&	6.41		& 45.9	&	538			& 1746	&\B{9}  	&\B{6.0}	&	88.1		&	9.01
		\end{tabular}
	\caption{Design relating to Villard de Honnecourt. 
	Given~$L_b$	and~$l$.}
	\label{tab:SketchBook-II}
\end{table}
The three cases with~$\ti=25^0$ or~$40^0$ and~$F=1/2$ were calculated 
\textit{a priori}.
The results illustrate the variation in going from the ideal~$F$-value of~1 
to the more realistic~2/3 or~1/2, and also the consequences when~$\ti$ is
varied from~$30^0$ to~$25^0$ or~$40^0$.
The variations are not very dramatic and the efficiencies are larger than~90\%
in all cases except when~$F=1/2$ and~$\en$ dips just below~90\%.

The large masses of the counterweights in~\tref{tab:SketchBook-II} can be 
reduced to the estimated maximum of~30 tons by relatively small reductions 
of the linear dimensions.
According to~\eref{eq:ExtraScalings} one finds for small 
changes~$\D R/R=\D L_i/L_i\simeq1/4\cdot\D M/M$ 
or about~8\% in the worst case, where~$R$ is reduced by~22m to~$275$m, 
but~$T$ drops strongly to~102kJ.
The reduction can also be implemented
by variation of the allowed maximum bending~$\be_2$ of the beam.
Bending is treated as a perturbation so it is not found in the
equations for the inner movement in~\ref{app:EOM},
but only in the scaling relations~\eref{eq:ExtraScalings} for lengths, 
masses and energies.
Therefore, in case~$\be_2$ is varied while one linear dimension is kept constant 
(like $L_b$ in~\tref{tab:SketchBook-II}), all other linear dimensions are 
invariants, while all masses, the energy~$T$ and the beam cross 
section~$(\pi/4)D^2$ are inversely proportional to~$\be_2$.
The counterweights~$M$ in~\tref{tab:CPNC} are therefore reduced if~$\be_2$
is increased (or beam strength weakened for same quality of wood).
Only modest reductions are necessary to bring most masses below~30 tons, but~35\%
is required for the heaviest and this is accomplished when~$\be_2$ is increased 
from~1/300 to~1/200.
This leaves the range invariant at~297m while~$T$ is reduced to~572kJ, 
less dramatically than above. 
It may seem odd that a smaller~$M$ implies a larger bending.
However, bending depends only on bending force~$Mg$ and 
beam strength~$\I\propto D^4$ under the given conditions, and~$D^2$ is proportional 
to~$m_b$ and~$M$.
The bending, which is proportional to~$Mg/\I$, therefore varies like~$1/M$.
Similar considerations apply to variation of the material constants~$\rho_d$ 
and~$\M_e$, or even gravity~$g$ for a trebuchet on a special mission to the moon.
\section{Summery and results}
The parameter space for the type of trebuchet discussed here
is very large, but reduced significantly by reasonable selections of initial conditions
and length of the arm for the counterweight.
The six free parameters that remain are listed in~\eref{eq:FreeAbsParameters},
but further reduction is possible because the equations for the internal 
movement can be scaled.
This leaves the four dimensionless parameters~$(l,l_4,\mu,\mu_b)$ 
in~\eref{eq:4free}.
A quality function~$\Q$ that depends only on these is defined in~\eref{eq:Q}.
It does not have a maximum under variation of all four parameters,
but a desired capacity and limited bending of the beam imposes a constraint 
among the parameters that secures a maximum.
This was found for many capacities and selected initial beam angles and 
counterweight arms.
The result can be summarized by a set of constants and a simple analytical 
expression given in~\eref{eq:ZRT}.
The comprehensive lists in~\tref{tab:CPNC} allow the determination of absolute 
dimensions and masses of optimized engines over a broad range of capacities.
\Eref{eq:ZRT} can also be used to link parameters two-by-two as done in several 
instances.
This is facilitated by the methods of linear algebra.
Evaluation can in every case be done simply by hand or more effectively with 
a spreadsheet.
This is the key result of the present study.

Two competing branches of optimized design exist in special cases. 
Even so, it is difficult to determine optimized design without guidelines. 
We explain why and illustrate this by considering a proposed
standard set of parameters to represent a large trebuchet.
It works but is far from optimal according to the present criteria.

A number of optimized design of archaeological interest are derived, and a 
clear trend is seen when the optimized dimensions
are compared with the ones attributed to historical trebuchets:
The optimized engines are smaller and the counterweights heavier.
Internal forces are also reduced and efficiencies increased.
\appendix
\section{Scaled equations of internal movement}\label{app:EOM}
Scaled time is~$\tau=(g/L_2)^{1/2}t$ and dot-notation is used for 
differentiation with respect to~$\tau$.
Scaled lengths~($l$,~$l_4$ and~$L$) and masses~($\mu$ and~$\mu_b$) are 
given in~\eref{eq:Design}.
\subsection{Phase I. Projectile in trough.}
The coupled differential equations for the motion of beam and counterweight are
\begin{eqnarray}
	&
	\left(\mu l^2f(\ta)^2+\frac{\mu_b}{3}\left(l^2-l+1\right)+1\right)\ddot\ta-
	L\cos(\ta-\psi)\ddot\psi  											\nonumber\\
	&+
	\mu l^2f(\ta)\frac{df}{d\ta}\dot\ta^2-
	L\sin(\ta-\psi)\dot\psi^2 											\nonumber\\
	&+
	\left(\frac{\mu_b}{2}\left(l-1\right)-1\right)\sin\ta=0
	\hspace{100pt}\mathrm{and}																							\label{eq:EOM_I_1}\\
	&L\ddot\psi-\cos(\ta-\psi)\ddot\ta+\sin(\ta-\psi)\dot\ta^2+\sin\psi=0~.
	\label{eq:EOM_I_2}
\end{eqnarray}
The sling angle~$\phi$ satisfies the geometric relation
\begin{eqnarray}\label{eq:EOM_I_3}
	\cos\ta+l_4\sin\phi=\cos\ta_i~,
\end{eqnarray}
where~$\ta_i$ is the initial beam angle, and  the
functions~$f$ and~$df/d\ta$ in~\ref{eq:EOM_I_1} are
\begin{eqnarray}\nonumber
	f(\ta)=
	\frac{\cos(\ta-\phi)}{\cos\phi} 
	\quad\mathrm{and}\quad
	\frac{df}{d\ta}=
	-\frac{\sin(\ta-\phi)}{\cos\phi}+
	\frac{\sin^2\ta}{l_4\cos^3\phi}~.
\end{eqnarray}
\subsection{Phase II. Projectile lifted from trough.}
The three equations for the motion of beam, counterweight and 
sling are equation~\eref{eq:EOM_I_2},
\begin{eqnarray}
	&
	\left(\mu l^2+\frac{\mu_b}{3}\left(l^2-l+1\right)+1\right)\ddot\ta-
	L\cos(\ta-\psi)\ddot\psi- 
	\mu l^2l_4\sin(\ta-\phi)\ddot\phi									\nonumber\\
	&-
	L\sin(\ta-\psi)\dot\psi^2+
	\mu l^2l_4\cos(\ta-\phi)\dot\phi^2 					\nonumber\\
	&+
	\left(\mu l+\frac{\mu_b}{2}\left(l-1\right)-1\right)\sin\ta=0
	\hspace{100pt}\mathrm{and}																							\label{eq:EOM_II_1}\\
	&ll_4\ddot\phi-l\sin(\ta-\phi)\ddot\ta-l\cos(\ta-\phi)\dot\ta^2-\cos\phi=0~.
	\label{eq:EOM_II_3}
\end{eqnarray}
\subsection{Phase III. Projectile released.}
The equations for the motion of beam and counterweight 
are~\eref{eq:EOM_II_1} with~$\mu=0$ and~\eref{eq:EOM_I_2}.
\section{Maximizing efficiency~$\en$}\label{app:ME}
The function~$\Q_\en$ defined by
\begin{eqnarray}\nonumber
	\Q_\en=\frac{\trho\hspace{2pt}^0\mu^0\en}{\F^0}=\en~,
\end{eqnarray}
has the form of a quality function, but does not qualify as such,
because it does not have a maximum under variation of~$(l,l_4,\mu_b)$ 
at a given capacity~$(R,T)$ and allowed bending~$\be_2$.
However, attempts at optimization reveals two maxima of~$\Q_\en=\en$
under variation of~$(l_4,\mu_b)$ with constant~$l$.
For high values of~$l$, one maximum is found at~$l_4<1$, the other 
at~$l_4>1$ and both are increasing functions of~$l$ that appear to 
converge towards~1.

\Tref{tab:SketchBook-I} with parameters relating to Villard de Honnecourt
shows optimized engines with the capacity~$R=450$m and~$T=232$kJ. 
Depending on the quality function used for the optimization, the
efficiency ranges up to~$\en=97.5\%$ for practical design parameters.
Calculations with the same capacity and~$\Q=\Q_\en$ show an increase 
of~$\en$ above~99.5\% at~$l=16.2$ on the smaller-$l_4$ branch 
and at~$l=58.5$ on the larger-$l_4$ branch.
These high values of~$l$ and the given capacity result in absolute 
parameters that can not be used in practical engines 
(see~\tref{tab:Large_Eff}), 
but the dynamics is interesting from a formal point of view.
\begin{table}[tbh]\footnotesize                                         
	\centering
		\begin{tabular}{cc|cccccc|cccc} 
			$l$	& $\en$ &$L_1$	&$L_2$	& $L_4$	& $M$		& $m$		&$m_b$	&	$\ta$			&	$\psi$		&	$\phi$		&	$A$				\\
			   	&  \%		& m			& m			& m			& ton		& kg		& kg 		& $-180^0$	&       		& $-180^0$	&						\\\hline
			17	& 99.53 & 0.800 & 0.047	& 0.673 & 270 	&	104   & 14.7	& $-1.95^0$ & $-1.98^0$	& $45.1^0$	& $44.9^0$	\\
			59	& 99.50 & 14.9  & 0.25 	& 18.2	& 71.5 	& 93.5	& 729		& $-1.22^0$	& $-0.91^0$	& $47.6^0$	& $42.6^0$			
		\end{tabular}
	\caption{	Design with high efficiencies~$\en$.
						$F=1$, $\ta_i=30^0$, $R=450$m and~$T=232$kJ.
						Angles of beam~$\ta$, counterweight~$\psi$ and sling~$\phi$ at release.
						$A$: initial climb of projectile.
						First row: Smaller-$l_4$ branch.
						Second row: Larger-$l_4$ branch.}
	\label{tab:Large_Eff}
\end{table}

As~$l$ is increased, both branches show decreasing relative projectile and 
beam masses which ensures an initial almost free fall of the counterweight. 
However, the fall must stop eventually and because of the one-sided focus on 
high efficiency, the relative parameters~$(l_4,\mu_b)$ are adjusted such that 
the descending motion is transformed smoothly at the last moment into an 
oscillatory motion of small amplitude.
 
This is seen most clearly in the limit~$l\rightarrow\infty$ on the smaller-$l_4$ 
branch where one of the two equations for the phase III motion takes the form
\begin{eqnarray}\nonumber
	\ddot\ta-L\cos(\ta-\psi)\ddot\psi-L\sin(\ta-\psi)\dot\psi^2-\sin\ta=0~,
\end{eqnarray}
while the other is unchanged.
The two equations are solved by
\begin{eqnarray}\nonumber
	(L+1)\ddot\psi+\sin\psi=0
	\quad\mathrm{and}\quad
	\ta-\psi=\pi
	~,
\end{eqnarray}
which for small amplitudes are harmonic oscillations.
These have dimensionless frequency~$\{1/(L+1)\}^{1/2}$
or in absolute terms~$\{g/(L_2+L_3)\}^{1/2}$.
The counterweight and beam thus swing in phase relative to the pivot as a 
quasi-solid body with pendulum length~$L_2+L_3$.
The transition from free fall of the counterweight to harmonic motion takes 
place over a very short period of time.
Counterweight and beam slow down during this period without ever stopping and go 
into the common oscillating mode such that the combined motion is towards the 
equilibrium point.
The reaction force from the fulcrum rises to a very high level during 
this transition.

A closer inspection of the motion for large but finite~$l$ reveals an
internal oscillation of high frequency on top of the harmonic component. 
The angle~$\ta-\psi$ oscillates around~$\pi$ with very small amplitude
as though the quasi-solid rings.
This is understood by including small terms as perturbations.
When the beam motion is not totally forced by the heavy counterweight it 
depends on the beam's moment of inertia relative to the pivot, but the torque on 
the beam is still dominated by the heavy counterweight of scaled mass~1.
The scaled moment of inertia is~$\I_P=1/3\cdot\mu_b\left(l^2-l+1\right)$ so
the frequency of small oscillations is~$\omega\simeq\I_P^{~-1/2}$ and a closer
analysis gives the more precise expression~$\omega=\{L/(L+1)\cdot\I_P\}^{-1/2}$.

For~$l=17$ in~\tref{tab:Large_Eff}, the reaction force at the culcrum reaches 
nearly~$10\hspace{1pt}Mg$, the tension of the sling rises to~$350\hspace{2pt}mg$ 
shortly before release, and the internal oscillations have~$f=42.2$Hz~. 
Further, as seen in~\tref{tab:Large_Eff} the configuration
at release is almost vertical, and the projectile is ejected from 
a low position~$\simeq2$m and at an initial climb close to~$45^0$.
Note also that the~270 ton counterweight falls only~$\simeq9$cm ($\simeq2L_2$).

The motion on the larger-$l_4$ branch is more complicated because the
mass of the beam never ceases to matter ($\mu_bl$ rises with~$l$).
The final oscillations of beam and counterweight are in phase with 
frequency~$\simeq\{g/(L_2+L_3)\}^{1/2}$ as before, 
but with small internal oscillations of a similar frequency 
(more like a flexible body), and the 
transition to the oscillating mode is towards maximum amplitude where the 
counterweight comes to rest before the final oscillations start.
This transition to harmonic motion is also much softer than the previous:
The reaction force from the bearings rises only a little above~$Mg$, 
and the sling tension stays below~$20\hspace{2pt}mg$. 
The configuration at release is again almost vertical, 
the projectile is ejected from a high position~$\simeq40$m at a slightly smaller 
initial climb~$42.6^0$, and the~71 ton counterweight now falls about~1/2~m. 


\section*{References}
\begin{thebibliography}{10}
%
	
\bibitem{ref:PVH} 
	Hansen, P. V., Acta Archaeologica, \textbf{63}, (1992), pp189-268

\bibitem{ref:Chev_et_al} 
	Chevedden, P. E., Eigenbrod, L., Foley, V., and Soedel, W. (1995),
	Sci. Am. \textbf{273}, pp66-71.

\bibitem{ref:Chev} 
	Chevedden, P. E.  (2000), 
	Dumbarton Oaks Papers, 54. Washington D.C.
	
\bibitem{ref:MSF2} 
	Fulton M.~S., Artillery in the Era of the Crusades: 
	Siege Warfare and the Development of Trebuchet Technology.
	History of Warfare, vol. 122, BRILL (2018).
	
\bibitem{ref:MSF1} 
	Fulton M. S., Chrissis N. G., Phillips J. and Kedar B. Z. 
	Crusades, (2017), \textbf{16}, pp33-53. 

\bibitem{ref:Galan} 
	Gal\'an M. R. and Gal\'an M. B.,
	Archaeometry 62, 5 (2020) 904–916

\bibitem{ref:TS} 
	Saimre T., Estonian Journal of Archaeology, (2006), \textbf{10}, 1, pp61-80.

\bibitem{ref:Denny} 
	Denny M., 2005 Eur. J. Phys. 26 561
	
\bibitem{ref:GG} 
	Goodno B. J. and Gere J. M. Mechanics of materials, 2009.
	
\bibitem{ref:VdH} 
Facsimile of the Sketch-Book of Villars de Honnecourt,  
J. B. A. Lassus and J. Quicherat, 
trans. and ed. R. Willis (London, 1859);

 
\end{thebibliography}
\end{document}